\documentclass[journal]{IEEEtran}
\usepackage{booktabs} 
\usepackage[utf8]{inputenc}
\usepackage{multirow}
\usepackage{graphicx}
\usepackage{epstopdf}
\usepackage{subfig}
\usepackage{subfloat}
\usepackage{verbatim}
\usepackage{cite}
\usepackage{color}
\usepackage{comment}
\usepackage{amssymb}
\usepackage{textcomp}
\usepackage{xcolor}
\usepackage{diagbox}
\usepackage{amsmath}
\usepackage{color} 
\usepackage{fancyhdr}

\usepackage{amsfonts} 
\usepackage{nomencl}
\usepackage{amsthm}

\usepackage{enumitem}
\usepackage{float}
\usepackage{slashbox}
\usepackage{graphicx}
\usepackage[ruled,linesnumbered]{algorithm2e}

\usepackage{tikz}
\newcommand*{\circled}[1]{\lower.7ex\hbox{\tikz\draw (0pt, 0pt)%
    circle (.5em) node {\makebox[1em][c]{\small #1}};}}

\def\lbrack{[}
\def\rbrack{]}

\begin{document}

\title{LESSON: Multi-Label Adversarial False Data Injection Attack for Deep Learning Locational Detection}

\author{Jiwei~Tian, Chao~Shen, Buhong~Wang, Xiaofang~Xia, Meng~Zhang, Chenhao~Lin, and Qian~Li 
	\thanks{This work is supported by the National Key Research and Development Program of China (2021YFB3100700), the National Natural Science Foundation of China (62161160337, 62132011, U21B2018, U20A20177, 62376210, 62206217), the Shaanxi Province Key Industry Innovation Program (2023-ZDLGY-38, 2021ZDLGY01-02),  the China Postdoctoral Science Foundation (2022M722530, 2023T160512), the Shaanxi Province Postdoctoral Research Funding Project (2023BSHYDZZ20), and the Fundamental Research Funds for the Central Universities (xzy012022082). J. Tian,  C. Shen,  M. Zhang, C. Lin and Q. Li are with the Ministry of Education Key Laboratory for Intelligent Networks and Network Security, School of Cyber Science and Engineering, Xi’an Jiaotong University, Xi’an, China (e-mail: tianjiwei2016@163.com; chaoshen@mail.xjtu.edu.cn; mengzhang2009@xjtu.edu.cn; linchenhao@xjtu.edu.cn; qianlix@xjtu.edu.cn). J. Tian is also with Air Traffic Control and Navigation College, Air Force Engineering University, Xi’an, China. Buhong Wang is with Information and Navigation College, Air Force Engineering University, Xi’an, China (e-mail: hongwks@aliyun.com). X. Xia is with the School of Computer Science and Technology, Xidian University, Xi’an, China. (e-mail: xiaofangxia89@gmail.com). (\textit{Corresponding author: Chao Shen.})}
}

\markboth{IEEE Transactions on Dependable and Secure Computing}%
{Shell \MakeLowercase{\textit{et al.}}: Bare Demo of IEEEtran.cls for IEEE Journals}

\maketitle

\begin{abstract}
Deep learning methods can not only detect false data injection attacks (FDIA) but also locate attacks of FDIA. Although  adversarial false data injection attacks (AFDIA) based on deep learning vulnerabilities have been studied in the field of single-label FDIA detection, the adversarial attack and defense against multi-label FDIA locational detection are still not involved. To bridge this gap, this paper first explores the multi-label adversarial example attacks against multi-label FDIA locational detectors and proposes a general multi-label adversarial attack framework, namely muLti-labEl adverSarial falSe data injectiON attack (LESSON). The proposed LESSON attack framework includes three key designs, namely Perturbing State Variables,  Tailored Loss Function Design, and Change of Variables, which can help find suitable multi-label adversarial perturbations within the physical constraints to circumvent both Bad Data Detection (BDD) and Neural Attack Location (NAL). Four typical LESSON attacks based on the proposed framework and two dimensions of attack objectives are examined, and the experimental results demonstrate the effectiveness of the proposed attack framework, posing serious and pressing security concerns in smart grids.

\end{abstract}

\begin{IEEEkeywords}
Adversarial example, bad data detection, deep learning, false data injection, multi-label learning, state estimation.
\end{IEEEkeywords}


\IEEEpeerreviewmaketitle
\makenomenclature
\renewcommand{\nomname}{NOMENCLATURE}
\mbox{} 
\nomenclature[01]{$m$}{The number of sensor measurements.}
\nomenclature[02]{$n$}{The number of state variables.}
\nomenclature[03]{$\boldsymbol{z}$}{A vector of measurements.}
\nomenclature[04]{$\boldsymbol{H}$}{A  Jacobian matrix.}
\nomenclature[05]{$\boldsymbol{x}$}{A vector of state variables.}
\nomenclature[06]{$\boldsymbol{e}$}{A  vector of measurement errors.}
\nomenclature[07]{$\boldsymbol{\hat{x}}$}{A  vector of estimated state variables.}
\nomenclature[08]{$\boldsymbol{R}$}{A  covariance matrix of $\boldsymbol{e}$.}
\nomenclature[09]{$\boldsymbol{a}$}{An attack vector.}
\nomenclature[10]{$\boldsymbol{c}$}{A  vector of malicious errors.}
\nomenclature[11]{$\tau$}{The detection threshold.}
\nomenclature[12]{$f(\cdot)$}{A function defined by a  neural network (NN).}
\nomenclature[13]{${\zeta(\cdot)}$}{A function computed before the softmax in a NN.}
\nomenclature[14]{$\boldsymbol{y}$}{An output vector of a NN.}
\nomenclature[15]{${\hat{y}}_{\boldsymbol{z}}$}{A predicted label of $\boldsymbol{z}$.}
\nomenclature[16]{$\boldsymbol{z}{'}$}{A vector of an adversarial example.}
\nomenclature[17]{$\boldsymbol{\vartheta}_{\boldsymbol{c}}$}{An adversarial perturbation vector.}
\nomenclature[18]{$\boldsymbol{\vartheta}$}{An adversarial perturbation matrix.}
\nomenclature[19]{$\Pi_{\boldsymbol{c}}$}{The indicator vector of non-zero elements of $\boldsymbol{c}$.}
\nomenclature[20]{$\sigma_i$}{The standard deviation of the $i$-th meter.}

\printnomenclature

\section*{Nomenclature}
\addcontentsline{toc}{section}{Nomenclature}
\begin{IEEEdescription}[\IEEEusemathlabelsep\IEEEsetlabelwidth{$V_1,V_2,V_3,V_3$}]
	\item[Acronyms]
	
	\item[AFDIA] Adversarial False Data Injection Attack
	\item[AML] Adversarial Machine Learning
	\item[BDD] Bad Data Detection
	\item[CNN] Convolutional Neural Network
	\item[DNN] Deep Neural Network
	\item[FDIA] False Data Injection Attack
	\item[LESSON] muLti-labEl adverSarial falSe data injectiON attack
	\item[NAD] Neural Attack Detection
	\item[NAL] Neural Attack Location
	\item[NN] Neural Network
	\item[WLS] Weighted Least Squares

	\item[]
	\item[Symbols]
	\item[$\boldsymbol{z}$] A vector of measurements
	\item[$\boldsymbol{H}$] A Jacobian matrix for state estimation
	\item[$\boldsymbol{x}$] A vector of state variables
	\item[$\boldsymbol{e}$] A  vector of measurement errors
	\item[$\boldsymbol{\hat{x}}$] A  vector of estimated state variables
	\item[$\boldsymbol{R}$] A  covariance matrix of $\boldsymbol{e}$
	\item[$\sigma_i$] The standard deviation of the $i$-th meter
	\item[$\tau$] The BDD detection threshold
	\item[$\boldsymbol{z_a}$] A vector of measurements after FDIA
	\item[$\boldsymbol{a}$] FDIA vector
	\item[$\boldsymbol{c}$] A  vector of malicious errors
	\item[$\boldsymbol{z_f}$] A vector of measurements after AFDIA
	\item[$f(\cdot)$] A function defined by a NN
	\item[$\zeta(\cdot)$] A function computed before Softmax of $f(\cdot)$
	\item[$\omega(\cdot)$] Softmax function
	\item[$\boldsymbol{y}$] An output vector of a NN
	\item[${\hat{y}}_{\boldsymbol{z}}$] A predicted label of $\boldsymbol{z}$
	\item[$\boldsymbol{\vartheta}$] A perturbation vector added to measurements

	\item[$\varTheta(\cdot)$] A NN multi-label classifier
	\item[$\varPsi(\cdot)$]  The multi-label predictor of $\varTheta$
	\item[$\varUpsilon(\cdot)$] A function computed before Sigmoid of $\varPsi(\cdot)$
	
	\item[$\boldsymbol{\zeta}$] A perturbation added to state variables
	
	\item[$\mu$] A predetermined state perturbation range
	\item[$\boldsymbol{\varpi}$] The change-of-variables for $\boldsymbol{\zeta}$
	
	\item[$\mathbf{I}$] A column vector of the same size as $\boldsymbol{c}$
	
	\item[]
	\item[Subscripts]
	\item[m] Number of meter measurements
	\item[n] Number of state variables

\end{IEEEdescription}

\section{Introduction}

\subsection{Background}

\IEEEPARstart  {F}{alse} data injection attacks (FDIA) have received substantial attention due to the stealthy characteristics and potential destructive effects in smart grids, such as line overloads, load shedding, cascading failures and blackout \cite{8293852,9764806,9815319,9676996,9563211,9141329}. Therefore, various of countermeasures have been proposed to thwart the FDIA threat, among which the detection methods based on artificial intelligence show excellent performance \cite{8887286}. Moreover, deep learning methods can not only detect attacks but also locate attacks \cite{9049087, tiis:24681, Mukherjee2022, MUKHERJEE2022100702, en15145312,Qin2022,9559412,HAN2023103016} to better deal with the threat of FDIA. However, current machine learning models proposed in power grids are vulnerable to various attacks \cite{chen2018machine,9642059,10003611}.  As presented in \cite{9622117,9695995,li2020conaml}, deep learning-based FDIA detectors are highly vulnerable to adversarial example attacks, leading to a more advanced and intelligent stealthy attack, called adversarial false data injection attack (AFDIA). AFDIA applies related laws of conventional false data injection attacks and adversarial example attacks to circumvent both the Bad Data Detection (BDD) and Neural Attack Detection (NAD) in Smart Grids. In this way, it can change the state of the grid without being detected to achieve specific attack objectives. In order to deal with this new attack threat, it is imperative to fully explore its attack methods and characteristics, so as to lay the foundation for designing effective defense measures.

\subsection{Related Work}
\textbf{Deep Learning based FDIA Detection and Localization:} Deep learning mainly uses deep neural networks (DNN) for training and prediction and has achieved remarkable results and progress in many fields, such as image recognition, speech recognition, natural language processing, and so on. Due to the powerful performance and excellent ability of deep learning, many methods based on DNN have been proposed to detect FDIA \cite{8791598}. The deep learning-based identification (DLBI) scheme in \cite{7926429} employs the Conditional Deep Belief Network (CDBN) to recognize the high-dimensional temporal features of the FDIA. Recently, a privacy-preserving collaborative learning method considering the local spatiotemporal relationship of measurements is presented to detect FDIA in AC-model power systems \cite{9508811}. Furthermore, a global spatiotemporal deep network, PowerFDNet, is proposed in \cite{9861714} to improve the detection performance of FDIA. Besides, considering system line parameters modeling errors, a deep transfer learning-based method is explored in \cite{2021arXiv210406307X} to detect the FDIA. Considering the privacy preservation issue, a FDIA detection method leveraging secure federated learning is proposed in \cite{9878267}.

Moreover, deep learning techniques are also used to locate false data injection attacks, showing the prospect of tracing the attack location to better deal with the FDIA threat. In \cite{9049087} and \cite{tiis:24681}, convolutional neural network (CNN) based locational detection methods are investigated to detect the exact locations of FDIA. References \cite{Mukherjee2022} and \cite{MUKHERJEE2022100702} emphasized FDIA detection as a multi-label classification problem and compared several deep learning models with conventional machine learning models to explore the best multi-label classifier for FDIA location. In \cite{en15145312}, a multivariate-based multi-label locational detection (MMLD) method is investigated to detect the FDIA presence and identify the attack locations, thus increasing the locational detection performance. To deal with the threat of coordinated state‑and‑topology FDIA, a multi-modal learning model is also presented to detect and locate this type of FDIA \cite{Qin2022}. In addition, using the graph topology of power systems and the spatial correlations of data \cite{9559412}, a graph neural network (GNN) based model is proposed to jointly detect and locate the FDIA in power grids. However, as pointed out in \cite{HAN2023103016}, the data generated by the sensors of branches and the connections between the branches are ignored in \cite{9559412}, which results in a relatively high false positive rate. To tackle the issue, based on graph data modeling and graph deep learning, a novel FDIA localization detection method is explored in \cite{HAN2023103016}, fully mining the correlation features between data components.

\textbf{Adversarial Example and AFDIA:} Although DNN models exhibit excellent performance, they are extremely vulnerable to adversarial example attacks \cite{9530723, 10092807, 9652053, 9965436, 9951062, zhu2023sgma}. The vulnerability of DNN-based renewable energy forecasting is studied and a novel cyberattack named adversarial learning attack is proposed in \cite{10255313}. A novel attack is introduced in \cite{10039065} to evade the existing global electricity theft detectors while stealing electricity. A systematic study is conducted in \cite {9914610} to evaluate the effectiveness of various adversarial example methods against NNN-based voltage stability assessment, and a novel lightweight mitigation strategy named robust online stability assessment (ROSA) is developed in \cite{10005029}. Meanwhile, the fragility of DNN also brings new security risks and threats to deep learning-based FDIA detectors \cite{WU2022108598}. In view of the significant differences between the power grid field and the image domain, the authors in \cite{li2020conaml} first proposed constrained adversarial machine learning (ConAML) to generate adversarial perturbations that meet the underlying constraints of the physical systems. Considering the final attack effect, the authors in \cite{9622117} introduced the joint adversarial example and FDIAs to investigate different attack scenarios for power system state estimation, where the new attack type AFDIA is investigated by taking advantage of the vulnerabilities of both BDD and NAD. Besides, the AFDIA method in \cite{9622117} is further extended to targeted AFDIA by the designed parallel optimization algorithm in \cite{9695995}. In addition, a practical and stealthy adversarial attack for generating adversarial perturbations based on the proposed unsupervised disentangled representation model is investigated in \cite{wu2021practical}. In response to the threat of AFDIA, the random input padding and adversarial training are respectively investigated in \cite{li2021adversarialresilient} and \cite{LIU2021102265}, showing a certain defensive effect. However, the above AFDIA attack and defense methods are both for single-label deep learning detectors. Due to the emergence of numerous multi-label deep learning detection models and their broad application prospects, it is urgent to explore their adversarial attack methods and defense strategies.

\textbf{Multi-label Adversarial Example:} Even in the image domain, the research on multi-label adversarial examples is much less than that on single-label adversarial examples. This is a nontrivial and challenging task because the number of positive labels is uncertain, multiple labels are usually not mutually exclusive, and some labels are often correlated. In this case, if an attacker tries to attack one or more specific labels, the confidence level of other non-attacking labels may also change. Thus, the attack methods for single-label classification cannot be directly applied to multi-label classification. The authors in \cite{8594975} proposed a universal adversarial attacking framework targeting at multi-label classifiers and conducted a preliminary analysis on the adversarial perturbations for DNN. The authors in \cite{9206614} presented a novel method that generates multi-label adversarial examples effectively by dealing with a linear programming problem, resulting in significantly smaller adversarial distortions. In \cite{9534067}, hiding all labels for multi-label images is investigated, and an empirical study of multi-label adversarial examples is presented.  In \cite{9710047}, a method to generate adversarial perturbations for top-k multi-label classifiers is proposed. The authors in \cite{2021Characterizing} assessed the attackability of multi-label learning systems and theoretically analyzed the bound of the expected worst-case risk. In \cite{9857594}, a differential evolution (DE) algorithm to effectively create multi-label adversarial perturbations is explored. Besides, domain knowledge and randomized smoothing are respectively employed in \cite{9661418} and \cite{JinyuanJia} to defend multi-label adversarial example attacks.

\begin{figure*}[htbp]
	\centerline{\includegraphics[scale=0.47]{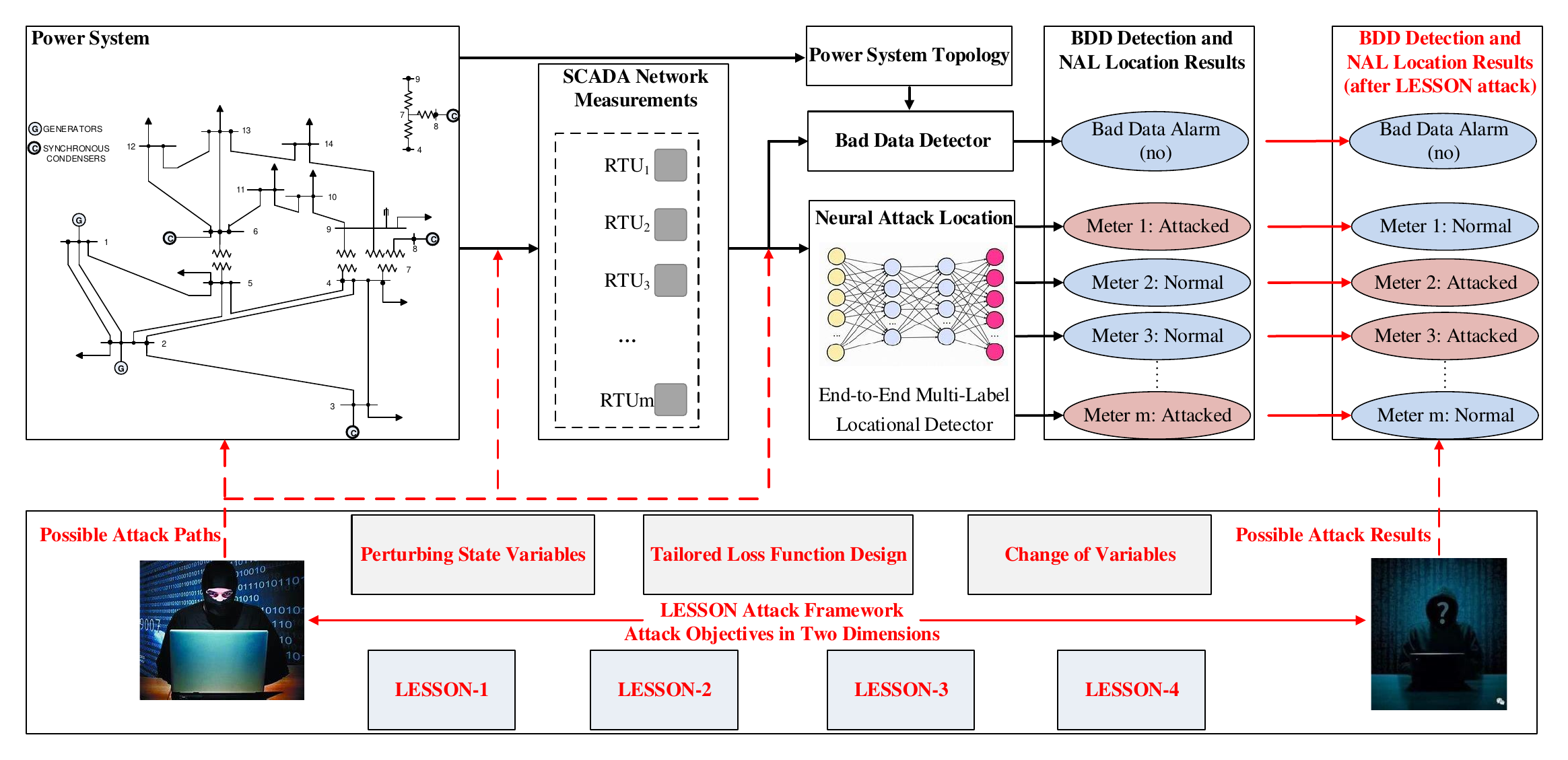}}
	\centering
	\caption{Multi-label deep learning FDIA locational detection and LESSON attack framework.}
	\label{attack_graph}
\end{figure*} 

\subsection{Contributions}

Considering the urgent need for multi-label adversarial example attacks and defenses, in this paper, we investigate the adversarial attack for multi-label FDIA locational detectors in the power grid field to provide a basis and reference for the subsequent defense strategies. To the best of our knowledge, this is the first work to explore multi-label adversarial example attacks in power systems, which can promote researchers in relevant fields to pay attention to and investigate this topic, so as to better deal with relevant security risks and threats.  Since effective defense methods for  multi-label adversarial examples have not been fully studied, the related characteristics of multi-label adversarial examples, multi-label learning and domain knowledge of power systems can provide strong support for subsequent defense measures research.

In summary, the main contributions  are described as follows:

$\bullet$ We propose a general and versatile multi-label adversarial attack framework for FDIA's locational detection, namely LESSON (Fig.~\ref{attack_graph}). The proposed LESSON attack framework includes three key designs, namely \textbf{Perturbing State Variables},  \textbf{Tailored Loss Function Design}, and \textbf{Change of Variables}, which can help find suitable multi-label adversarial perturbations within the physical constraints to circumvent both BDD and NAL's detection:
	\begin{enumerate}
	    \item The \textbf{Perturbing State Variables} can ensure that the generated multi-label adversarial perturbation meets the mutual relationship of physical quantities in power systems so that the corresponding LESSON attacks will not be detected by BDD. 
	    \item The \textbf{Tailored Loss Function Design} provides flexible loss functions that can be adjusted according to the attack objective, and many excellent optimizers such as Adam can be used to optimize the suitable multi-label adversarial perturbation.
 	    \item The \textbf{Change of Variables} can flexibly limit the range of perturbations to ensure that the generated multi-label adversarial perturbation does not violate the physical constraints of power systems, resulting in practical and effective multi-label adversarial example attacks. 
	\end{enumerate} 

$\bullet$ We explore four typical LESSON attacks (LESSON-1 to LESSON-4) based on the proposed attack framework and two dimensions of attack objectives. \textbf{Objective One} is concerned with the induced estimation error, which represents the impact of attackers' attempts on the state estimation results of power systems. \textbf{Objective Two} is related to the locational detection results of the multi-label FDIA detector for attacks, indicating the spoofing ability of the attacker to the multi-label FDIA location detector and the true concealment of the attacks. The four typical LESSON attacks based on the attack objectives of the two dimensions are very representative and can be used to analyze and evaluate the security risks faced by multi-label classifiers in power systems.

$\bullet$ We conducted extensive experimental analyses to investigate the proposed four LESSSON attacks, and analyzed the relevant influencing factors:   

\begin{enumerate}
	\item The results demonstrate that the proposed LEESON attack framework has a very good attack success rate. The success rate of all four LESSON attacks of the 118-bus system is 100$\%$ (even in the most difficult LESSSON-4 attack case), posing a huge threat to large-scale power systems. Besides, the most difficult LESSON-4 attack of the 14-bus and 30-bus systems still has a success rate of more than 60$\%$, and the success rate of most other LESSON cases is close to 100$\%$.  
	\item The initial FDIA attack scale has a certain impact on the final attack success rate: as the scale of FDIA attacks increases, larger adversarial perturbations are usually required to bypass detection, and the limitations of physical constraints make it difficult to find suitable adversarial perturbations, resulting in a decrease in attack success rate.
	\item With the increase of the FDIA attack scale, Adam's initial learning rate should also be appropriately increased to more effectively seek appropriate adversarial perturbations, so as to improve the attack success rate: for the case \textbf{Objective Two} \circled{2} (LESSON-2 and LESSSON-4), the original learning rate should be somewhat smaller, otherwise the success rate will quickly decrease to 0; for other cases, the chosen original learning rate cannot be too large or too small, which can be roughly in the range of [0.001,0.1] and adjusted according to the FDIA attack scale. 

\end{enumerate}

The remainder of the paper is organized as follows. In Section II, the preliminary materials are briefly introduced. In Section~III, the proposed LEESON attack framework is presented. In Section IV, the proposed four LESSSON attacks are evaluated and analyzed for a variety of case studies, followed by a conclusion in Section V.

\section{PRELIMINARIES}

\subsection{FDIA and AFDIA}

For the Direct-current (DC) state estimation, the model for an $n+1$ buses' power system with $m$ meters is as follows\footnote{Although DC state estimation approaches are used in this work, we think that AC state estimation models are also likely to have similar flaws. Therefore, in subsequent work, we will conduct a thorough investigation of AC state estimation methods.}:
\begin{equation}
\boldsymbol{z}=\boldsymbol{Hx}+\boldsymbol{e},
\label{eq1}
\end{equation}
where $\boldsymbol{z} \in \mathbb{R}^{m \times 1}$, $\boldsymbol{x} \in \mathbb{R}^{n \times 1}$ and $\boldsymbol{e}  \in \mathbb{R}^{m \times 1}$  stand for a meter measurements' vector, a state variable vector,  and a measurement errors' vector (noise), respectively.  The measurement Jacobian matrix is represented by $\boldsymbol{H} \in \mathbb{R}^{m\times n}$. The measurement errors are assumed to follow a zero mean and $\boldsymbol{R}$ covariance matrix's  Gaussian distribution, where $\boldsymbol{R}=\text{diag}(\sigma _{1}^{2},\sigma _{2}^{2},\cdots,\sigma _{m}^{2})$ and ${\boldsymbol{R}_{ii} \triangleq \sigma_{i}^{2}}$ denotes the variance of the $i$-th meter.

The Weighted-Least-Squares (WLS) state estimation, which is often employed, minimizes the following function \cite{Liu}
\begin{equation}
L(\boldsymbol{x}) =\sum_{i=1}^m{(\boldsymbol{z}_i-\boldsymbol{H}_i\boldsymbol{x}) ^2/\boldsymbol{R}_{ii}},
\label{wls}
\end{equation} where $\boldsymbol{H}_i$ represents the $i$-th row of $\boldsymbol{H}$.
Based on \eqref{wls}, the optimal state estimation result can be derived by
\begin{equation}
\label{x_estimation}
\boldsymbol{\hat{x}}=(\boldsymbol{H}^T\boldsymbol{R}^{-1}\boldsymbol{H}) ^{-1}\boldsymbol{H}^T\boldsymbol{R}^{-1}\boldsymbol{z},
\end{equation}
where $\boldsymbol{H}^T$ denotes the matrix $\boldsymbol{H}$'s transpose and $\boldsymbol{R}^{-1}$ denotes the matrix $\boldsymbol{R}$'s inverse. Erroneous measurements may be introduced for a variety of reasons including meter malfunction, communication failure, and cyber attacks. BDD just utilizes the residual $\boldsymbol{r}=\boldsymbol{z}-\boldsymbol{H\hat{x}}$ to identify bad data. $L(\boldsymbol{\hat{x}})$ is theoretically proved to follow the $\chi^2$ distribution by presuming that the meter errors $(\boldsymbol{z}_i-\boldsymbol{H}_i\boldsymbol{\hat{x}})/\sigma_{i}$ follow Gaussian distribution \cite{abur2004power}:
\begin{equation}
\label{ka}
\begin{split}
L(\hat{\boldsymbol{x}}) =\sum_{i=1}^m{ \Big (\frac{\boldsymbol{r}_i}{\sigma_i} \Big )^2}=\sum_{i=1}^m{\Big ( \frac{\boldsymbol{z}_i-\boldsymbol{H}_i\hat{\boldsymbol{x}}}{\sigma_{i}} \Big )}^2 \\
=\left( \boldsymbol{z}-\boldsymbol{H\hat{x}} \right)^{T} \boldsymbol{R}^{-1}\left( \boldsymbol{z}-\boldsymbol{H\hat{x}} \right). 
\end{split}
\end{equation}

$\boldsymbol{z}$ is recognized as a bad measurement vector if $L(\hat{\boldsymbol{x}}) \ge \tau$ ($\tau$ is the predefined threshold). If not, $\boldsymbol{z}$ is regarded as a normal data vector. Following a $\chi^2$ test with the desired level of significance (e.g., 99$\%$), the threshold $\tau > 0$ is predefined. Combining  Equations (\ref{x_estimation}) and (\ref{ka}), we can obtain 

\begin{equation}
	\label{ka_a}
	\begin{split}
	L(\boldsymbol{z}) =\left\{\boldsymbol{z}-\boldsymbol{H} \left( (\boldsymbol{H}^T\boldsymbol{R}^{-1}\boldsymbol{H}) ^{-1}\boldsymbol{H}^T\boldsymbol{R}^{-1}\boldsymbol{z} \right) \right\}^{T} \boldsymbol{R}^{-1} \\ \left\{\boldsymbol{z}-\boldsymbol{H} \left( (\boldsymbol{H}^T\boldsymbol{R}^{-1}\boldsymbol{H}) ^{-1}\boldsymbol{H}^T\boldsymbol{R}^{-1}\boldsymbol{z} \right) \right\}. 
	\end{split}
\end{equation}

However, as demonstrated in \cite{Liu}, an attacker can implement stealthy FDIA if the designed attack vector $\boldsymbol{a}$ meets the condition
\begin{equation}
\label{fdi}
\boldsymbol{a}=\boldsymbol{Hc},
\end{equation}
where $\boldsymbol{c}$  is an $n$-length non-zero column vector, representing the predetermined attack objective. The rationale is that 
\begin{equation}
\label{residue}
\boldsymbol{r}_a =\boldsymbol{z}_{\boldsymbol{a}}-\boldsymbol{H\hat{\boldsymbol{x}}}_{\boldsymbol{a}} 
=(\boldsymbol{z}+\boldsymbol{a})-\boldsymbol{H}(\hat{x}+\boldsymbol{c})=\boldsymbol{z}-\boldsymbol{H}\hat{x}
=\boldsymbol{r},
\end{equation}
where $\boldsymbol{z}_{\boldsymbol{a}}=\boldsymbol{z}+\boldsymbol{a}$, $\boldsymbol{r}_a$, $\hat{\boldsymbol{x}}_{\boldsymbol{a}}$ represent the tampered measurement, residual, state estimation result after FDIA, respectively. The Equation (\ref{residue}) means that the residual $\boldsymbol{r}_a$ under the FDIA is the same as the original residual $\boldsymbol{r}$. Then $L(\hat{\boldsymbol{x}}_a)=L(\hat{\boldsymbol{x}})$ can be derived. If $L(\hat{\boldsymbol{x}})$ meets the condition $L(\hat{\boldsymbol{x}}) \le \tau$, then $L(\hat{\boldsymbol{x}}_a)$, which corresponds to $\boldsymbol{r}_a$, also meets the condition $L(\hat{\boldsymbol{x}}_a) \le \tau$, indicating that the FDIA is stealthy.

To address the threat of FDIA, the NAD module is used to detect whether FDIA occurs, where a normal sample's label is 0, and an attacked sample's label is 1. A function ${f(\boldsymbol{z}:\Psi)=\boldsymbol{y}}$ that receives a measurement vector $\boldsymbol{z}\in \mathbb{R}^{m \times 1}$ and outputs $
{\boldsymbol{y}\in \mathbb{R}^{\kappa \times 1}}$ ($\kappa=2$), where $\Psi$ denotes the network parameters. In general, the Softmax function $\omega(\cdot)$ is often used for the network's output of classification tasks  \cite{gao2017properties}, and the well-trained NAD model $f$ classifies the sample by
\begin{equation}
    \hat{y}_{\boldsymbol{z}}=\max \limits _{y_i} \{ y_i | y_i \in \{ y_1, y_2 , \ldots,y_\kappa \}: \boldsymbol{y} = f(\boldsymbol{z}:\Psi) \}, 
	\label{function}
\end{equation}
where ${y_i}$ stands for the $\boldsymbol{y}$'s $i$-th variable and indicates the likelihood that $\boldsymbol{z}$ will be assigned to the $i$-th label. The output from the function $\zeta(\boldsymbol{z})$ is passed to the Softmax function $\omega(\cdot)$ by
\begin{equation}
f(\boldsymbol{z}:\Psi) ={\rm \omega}\left(\zeta\left( \boldsymbol{z} \right) \right) =\boldsymbol{y},
\label{pre_softmax}
\end{equation}
where $\zeta(\cdot)$ represents the output of ${f}$ before the Softmax fuction.

The following constrained optimization problem can be employed to generate an adversarial example $\boldsymbol{z}{'}$ of $\boldsymbol{z}$:
\begin{subequations}
	\label{uuuuu}
	\begin{align} 
	\label{OP:NOR}
	\min_{\boldsymbol{z}'} \quad  &\varphi( \boldsymbol{z}',\boldsymbol{z} )   \\
	\label{CON_N}
	s.t.\ \quad &\hat{y}_{\boldsymbol{z}'} \neq \hat{y}_{\boldsymbol{z}}. 
	\end{align}
\end{subequations}

Let $\boldsymbol{\vartheta} \triangleq \boldsymbol{z}'-\boldsymbol{z}$ represent the added perturbation. The distance function $\varphi( \boldsymbol{z}',\boldsymbol{z})=\lVert \boldsymbol{\vartheta} \rVert $ may be computed by $\ell_0$, $\ell_2$, or $\ell_\infty$ norm function. The aforementioned optimization problem (\ref{uuuuu}) minimizes the adversarial perturbation while causing models to misclassify the input under specific restrictions.

Adversarial false data injection attacks (AFDIA) proposed in \cite{9622117} try to deceive NAD to classify attacked/false samples as being legitimate/normal, where the key is to generate adversarial examples for NAD models. Since the well-trained NAD model will identify the measurement vector after FDIA ($\boldsymbol{z}_{\boldsymbol{a}}=\boldsymbol{z}+{\boldsymbol{a}}$) as attacked/false, a wise attacker needs to add an appropriate perturbation $\boldsymbol{\vartheta}$ to $\boldsymbol{z}_{\boldsymbol{a}}$ so that the final adversarial measurement vector after AFDIA ($\boldsymbol{z}_f=\boldsymbol{z}_{\boldsymbol{a}}+\boldsymbol{\vartheta}=\boldsymbol{z}+{\boldsymbol{a}}+\boldsymbol{\vartheta}$) can bypass the NAD's detection, where $\boldsymbol{a}{'} = {\boldsymbol{a}}+\boldsymbol{\vartheta}$ stands for the AFDIA's total attack vector. The above-mentioned procedure can be stated as
	\begin{subequations}
		\label{uuio}
		\begin{align} 
		\label{bbbb}
		\min_{\boldsymbol{\vartheta}} \quad &\varphi(\boldsymbol{z}_f,\boldsymbol{z}_{\boldsymbol{a}})   \\
		\label{cccc}
		s.t.\ \ \ \boldsymbol{z}_f=&\boldsymbol{z}_{\boldsymbol{a}}+\boldsymbol{\vartheta} \\
		\label{m111}
		\quad \quad \hat{y}_{\boldsymbol{z}_{\boldsymbol{a}}}&=1 \\
		\label{m222}
		\quad \quad \hat{y}_{\boldsymbol{z}_f}&=0
		\end{align}	
	\end{subequations}
where $\varphi(\boldsymbol{z}_f,\boldsymbol{z}_{\boldsymbol{a}})$ = $\lVert \boldsymbol{z}_f-\boldsymbol{z}_{\boldsymbol{a}} \rVert _2$. The constraints \eqref{m111} and \eqref{m222} imply that the  measurement vector after FDIA $\boldsymbol{z}_{\boldsymbol{a}}$ is classified as being attacked, and the final measurement vector after AFDIA $\boldsymbol{z}_f$ is identified as being legitimate.

\subsection{Multi-Label Learning and Multi-Label Attack Location}
Multi-label learning investigates the problem where each sample is represented by an instance while associated with a group of labels simultaneously \cite{6471714}. For the FDIA location problem, the multi-label learning process is stated as follows:
Assume that $ \boldsymbol{z}  \in \mathbb{R}^{m \times 1}$ is a real-time measurement vector, $ \boldsymbol{y}  \in \left\{ 0,1 \right\}^{m \times 1}$ is a label vector for $ \boldsymbol{z}$, where $ \boldsymbol{y}_j =1 $ or 0 means the $j$th label (meter/measurement) of $ \boldsymbol{y} $ is attacked or normal.  Let $u$ denote the number of training samples.  Then the sample matrix is $ \boldsymbol{Z}  \in \mathbb{R}^{u \times m}$, and the label matrix is $ \boldsymbol{Y}  \in \left\{ 0,1 \right\}^{u \times m}$.  Let $\varTheta:R^m \rightarrow \left\{ 0,1 \right\} ^m$ represent the multi-label classifier. The process of training a multi-label classifier for real-time FDIA locational detection is presented in Algorithm \ref{CNN_tarin}. Moreover, $\varPsi:R^m \rightarrow R^m$ represents the multi-label predictor, and the predicted value can be seen as the relevance confidence. Both $\varTheta$ and $\varPsi$ can be decomposed as $m$ sub-functions, i.e., $\varTheta= \left\{ \varTheta_1,\cdots ,\varTheta_m \right\} $ and $\varPsi= \left\{ \varPsi_1,\cdots ,\varPsi_m \right\}$. Besides, $\varTheta$ can be derived from $\varPsi$ via thresholding functions. For example, $\varTheta(\boldsymbol{z})_{j} = \lbrack \varPsi(\boldsymbol{z})_{j} > \chi \rbrack$ uses a thresholding function and outputs 1 if the predicted value is higher than the threshold $\chi$. $\lbrack \pi \rbrack$ returns 1 if predicate $\pi$ holds, and 0 otherwise.

\begin{algorithm}
	\SetAlgoLined
	
	\textbf{Training a multi-label classifier for FDIA detection}\\
	\textbf{Input:} \\
	training dataset $D_{tr}=$\{$\boldsymbol{Z} \in \mathbb{R}^{u \times m}$, $\boldsymbol{Y} \in \left\{ 0,1 \right\}^{u \times m}$\}, \\
	number of iterations used for training $N_{iter}$\\
	\textbf{Output:} A well-trained trained multi-label classifier $\varTheta$\\
	  
	\For{1,2,...,$N_{iter}$}{
	training the multi-label classifier $\varTheta$ using $D_{tr}$
	}
	  
	\textbf{Real-time FDIA locational detection by $\varTheta$}\\
	\textbf{Input:} \\
	well-trained multi-label classifier $\varTheta$, \\
	real-time measurement vector $\boldsymbol{z}  \in \mathbb{R}^{m \times 1}$\\
	\textbf{Output:} label prediction vector $\boldsymbol{y}  \in \left\{ 0,1 \right\}^{m \times 1}$ for $ \boldsymbol{z}$\\
	
	\For{$\boldsymbol{z}$}{
	$\boldsymbol{y} = \varTheta(\boldsymbol{z})$
	} 

	\caption{Training a multi-label classifier for real-time FDIA locational detection}
	\label{CNN_tarin}
\end{algorithm}

\section{LESSON: muLti-labEl adverSarial falSe data injectiON attack}

Based on the fragile characteristics of DNN and the relevant research on multi-label adversarial examples in the image domain, we extend AFDIA \cite{9622117}\cite{9695995} for single-label NAD classifiers to LESSON (muLti-labEl adverSarial falSe data injectiON attack) for multi-label NAL (Neural Attack Location) classifiers. In the following sections, we formulate and describe the designed LESSSON attack framework to address the problem.

\subsection{Attack Model and Objective}

The threat model considered in this work is as follows:
\begin{itemize}[leftmargin=*]
	\item  \textbf{Attack Condition}: The attacker is supposed to have knowledge of the power grid and NAL parameters. Actually, attackers can employ a variety of methods to acquire relevant information, such as insider threats \cite{baracaldo2013adaptive}, eavesdropping on information, and invading database systems \cite{9622117}. Although the assumption provides powerful attackers that may not always represent the actual situation, it enables us to evaluate the robustness and vulnerability in the worst case, thus providing an upper limit on the impact of attacks.
	\item \textbf{Attack Objective}: Both BDD and NAL methods are assumed to be used in power systems: BDD method is used to detect bad data, and NAL method is to locate conventional FDIA. The attacker has two objectives: \textbf{Objective One} is to tamper with power system state estimation results; \textbf{Objective Two} is to mislead the NAL's locational detection results. Note that, in order to consider a more comprehensive FDIA strategy, the attacker does not seek low sparsity FDIA\footnote{Low-sparsity FDIA seeks as few non-zero items in the attack vector $\boldsymbol{a}$ as possible, where the related target state variables must meet the specific relationship and are not random.}, so the relevant values of target state variables are arbitrary. 
	\item \textbf{Attack Cost}: We also assume that to achieve predefined initial attack objectives (perturbing certain state variables),  the relevant meters are controlled by the attacker. Specifically, if the attacker aims to disturb certain target state variables (nonzero items of $\boldsymbol{c}$), then the corresponding non-zero items (meters) of $\boldsymbol{a}$ by (\ref{fdi}) are controlled by the attacker \cite{9622117}. Besides, the attacker is assumed to have other resources to control other meters to add adversarial perturbations that can bypass NAL's detection \cite{9695995}.
\end{itemize}

\subsection{Perturbing State Variables for Multi-Label Adversarial Perturbations}

Based on the predetermined attack objective ($\boldsymbol{c}$), the attacker can use formula \eqref{fdi} to derive the required FDIA attack vector $\boldsymbol{a}$. A well-trained NAL model, however, is likely to be able to locate the FDIA. The crafty attacker can then implement additional adversarial attacks to evade the NAL's locational detection by exploiting the vulnerabilities in the NAL model.

Given an instance $\boldsymbol{z}$, let the abovementioned $\varTheta:R^m \rightarrow \left\{ 0,1 \right\} ^m$ be  a multi-label classifier satisfying $\varTheta(\boldsymbol{z})=\boldsymbol{y}$, where $\boldsymbol{y}$ is the prediction label vector of $\boldsymbol{z}$. Here, suppose $\varTheta$ correctly classify all labels of $\boldsymbol{z}$. Although it may not always be correct in practical situations, we can only cover the labels that $\varTheta$ correctly classifies $\boldsymbol{z}$ into  to easily achieve this. Besides, even for meters that are not correctly classified by the multi-label classifier, iterative solutions of adversarial perturbations below can still be performed because meters that already satisfies the attack objective do not generate additional gradient information for adversarial perturbations during the iteration.

Then the procedure to generate multi-label adversarial perturbations can be stated as
\begin{subequations}
		\label{uuioo}
		\begin{align} 
		\label{bbbbo}
		\min_{\boldsymbol{\vartheta}} \quad &\varphi(\boldsymbol{z}_f,\boldsymbol{z}_{\boldsymbol{a}})   \\
		\label{cccco}
		s.t.\ \ \ \boldsymbol{z}_f&=\boldsymbol{z}_{\boldsymbol{a}}+\boldsymbol{\vartheta} \\
		\label{vvv0}
		L(&\hat{\boldsymbol{z}_f}) \le \tau \\
		\label{m111o}
		\quad \quad \varTheta(\boldsymbol{z}_f)_i&=1, i\in A \\
		\label{m222o}
		\quad \quad \varTheta(\boldsymbol{z}_f)_i&=0, i\in B
		\end{align}	
\end{subequations}
where $\varphi(\boldsymbol{z}_f,\boldsymbol{z}_{\boldsymbol{a}}) = \lVert \boldsymbol{z}_f-\boldsymbol{z}_{\boldsymbol{a}} \rVert= \lVert \boldsymbol{\vartheta} \rVert$, and $\boldsymbol{z}_{\boldsymbol{a}}$ is the measurement vector after FDIA. The constraint \eqref{vvv0} implies that the final measurement vector $\boldsymbol{z}_f$ is categorized as being normal by BDD. The constraints \eqref{m111o} and \eqref{m222o} represent the misleading prediction results for the NAL, where sets A (predicated as 1) and B (predicated as 0) denote the corresponding meters in power systems. 

According to the relevant results in \cite{9622117}, it is difficult to solve (\ref{uuioo}) directly. Based on the relevant ideas in \cite{9622117}, we also consider generating adversarial perturbations through disturbing state variables. In this way, the generated adversarial perturbations will never be detected by BDD, which means that the constraint (\ref{vvv0}) can be removed. The relevant proofs and results can refer to \cite{9622117}, which are omitted here. Then, the optimization problem of multi-label adversarial perturbations can be expressed as
\begin{subequations}
	\label{uuioo2}
	\begin{align} 
	\label{bbbbo2}
	\min_{\boldsymbol{\zeta}}& \quad  \lVert \boldsymbol{\zeta} \rVert    \\
	s.t.\ \ \ \boldsymbol{z}_f&= \boldsymbol{z}_{\boldsymbol{a}}+\boldsymbol{H} \boldsymbol{\zeta} \\
	\label{m111o2}
	\quad \quad \varTheta(\boldsymbol{z}_f)&_i=1, i\in A \\
	\label{m222o2}
	\quad \quad \varTheta(\boldsymbol{z}_f)&_i=0, i\in B
	\end{align}	
\end{subequations}
where $\boldsymbol{\zeta}$ represents the perturbation added to state variables, and $\boldsymbol{H} \boldsymbol{\zeta}$ is the final adversarial perturbation for the FDIA measurement vector $\boldsymbol{z}_{\boldsymbol{a}}$.

\subsection{Tailored Loss Function Design for Multi-Label Adversarial Perturbations}
For FDIA's location classifier NAL, $\varPsi(\boldsymbol{z}_f)_{j} = Sigmoid(\varUpsilon(\boldsymbol{z}_f)_{j})$ , where the Sigmoid activation function is applied to the nodes of the output layer ($\varUpsilon(\boldsymbol{z}_f)$). Then $\chi$ is set as 0.5 as in \cite{8594975} for the abovementioned  $\varTheta(\boldsymbol{z}_f)_{j} = \lbrack \varPsi(\boldsymbol{z}_f)_{j} > \chi \rbrack$ thresholding function  to determine the labels of measurements. According to the properties of the Sigmoid function, we can know that $\varTheta(\boldsymbol{z}_f)_{j}=1$ is equivalent to $\varUpsilon(\boldsymbol{z}_f)$ greater than 0, and otherwise $\varTheta(\boldsymbol{z}_f)_{j}=0$.

Then, the constraints \eqref{m111o2} and \eqref{m222o2} can be transformed as follows:

\begin{subequations}
	\label{uuioo22}
	\begin{align} 
	\label{m111o22}
	\quad \quad \varTheta(\boldsymbol{z}_f)_i=1, i\in A \Leftrightarrow  \varUpsilon(\boldsymbol{z}_f)_i>0, i\in A \\
	\label{m222o22}
	\quad \quad \varTheta(\boldsymbol{z}_f)_i=0, i\in B \Leftrightarrow  \varUpsilon(\boldsymbol{z}_f)_i\le0, i\in B
	\end{align}	
\end{subequations}

Then, a straightforward way to solve the above problem is to convert the constraints to regularizers such as:

\begin{subequations}
	\label{uuiooa}
	\begin{align} 
	\label{bbbboa}
	\min_{\boldsymbol{\zeta}} \quad  \lVert \boldsymbol{\zeta} \rVert  + \lambda (\sum_{i=1}^A{\max \left( 0, -\varUpsilon(\boldsymbol{z}_f)_j \right)}&+\sum_{j=1}^B{\max \left( 0, \varUpsilon(\boldsymbol{z}_f)_j \right)})  \\
	s.t.\ \ \ \boldsymbol{z}_f= \boldsymbol{z}_{\boldsymbol{a}}+&\boldsymbol{H} \boldsymbol{\zeta}
	\end{align}	
\end{subequations}
where $\lambda$ is a trade-off penalty between the perturbation size ($\lVert \boldsymbol{\zeta} \rVert$) and attacking accuracy $(\sum_{i=1}^A{\max \left( 0, -\varUpsilon(\boldsymbol{z}_f)_j \right)}+\sum_{j=1}^B{\max \left( 0, \varUpsilon(\boldsymbol{z}_f)_j \right)})$.

\subsection{Change of Variables for Multi-Label Adversarial Perturbations}
Although the optimization problem (\ref{uuiooa}) can be used to generate multi-label adversarial perturbations, it indeed does not restrict the perturbation size, which may result in large and unrealistic adversarial perturbations. To alleviate this issue, we consider the change-of-variable technique as in \cite{9622117, 9695995}.

Define a new variable $\boldsymbol{\varpi}$, and let $\boldsymbol{\zeta} =\mu \cdot \tanh(\boldsymbol{\varpi})$. Since $-1 \le \tanh(\boldsymbol{\varpi}) \le 1 $, then $-\mu \le \boldsymbol{\zeta} \le \mu $, and the result ($\boldsymbol{\zeta}$) falls in the predetermined range. In this way, we can optimize over $\boldsymbol{\varpi}$ rather than optimizing over  $\boldsymbol{\zeta}$ by

\begin{subequations}
	\label{uuiooa7}
	\begin{align} 
	\label{bbbboa7}
	\min_{\boldsymbol{\varpi}} \quad  \sum_{i=1}^A{\max \left( 0, -\varUpsilon(\boldsymbol{z}_f)_j \right)}&+\sum_{j=1}^B{\max \left( 0, \varUpsilon(\boldsymbol{z}_f)_j \right)}  \\
	s.t.\ \ \ \boldsymbol{z}_f= \boldsymbol{z}_{\boldsymbol{a}}+&\boldsymbol{H} (\mu \cdot \tanh(\boldsymbol{\varpi})),
	\end{align}	
\end{subequations}
where $\mu$ represents the predetermined  state perturbation range. In the image domain, the image pixels are between [0,1] or [0, 255]. In the field of power systems, multi-label adversarial perturbations can be limited in the above way, such as setting $\mu$ as 1 or 0.5 \cite{9695995}.

With the change-of-variable technique described above, we can apply optimization methods without box limitations. Many optimization algorithms, such as stochastic gradient descent(SGD) and RMSProp, can be used to solve the problem. The adaptive optimization algorithm Adam is employed since it has better optimization performance. It is worth noting that the corresponding algorithm will be stopped if one feasible solution can be found. Therefore, creating multi-label adversarial perturbations will be effective.

\subsection{Typical Multi-Label Adversarial Perturbation Attacks}
The above optimization problem (\ref{uuiooa7}) provides a unified attack framework to generate required multi-label adversarial perturbations. The attacker can refine the corresponding parameters in (\ref{uuiooa7}) according to his objectives to generate appropriate adversarial perturbations. According to the attack objectives in two dimensions (\textbf{Objective One} and \textbf{Objective Two}), we mainly explore four typical attack types based on the designed LESSON attack framework. \textbf{Objective One} is the attack objective related to state estimation results, and \textbf{Objective Two} is the attack objective related to meter locational detection results. For \textbf{Objective One}, since perturbing state variables may affect the original induced estimation error\footnote{The original induced estimation error represents the specific error that the attacker initially intends to introduce to specific state variables, which is represented by the nonzero items/variables of $\boldsymbol{c}$ in Equation (\ref{fdi}).}, we divide the LESSON into two types. One is to keep the original induced estimation error unchanged, and the other does not necessarily meet such requirements. Obviously, keeping the original induced estimation error unchanged can accurately achieve specific attack targets, which is more difficult and can be seen as the \textbf{targeted} LESSON attack. In reality, the \textbf{targeted} LESSON attack is more important because it can accurately achieve specific attack targets, leading to more precise and destructive strikes to the power grid. For example, if an attacker attempts to introduce precise errors to specific buses to harm the power grid, the \textbf{targeted} LESSON attack can covertly complete this task without changing the original induced estimation error. For \textbf{Objective Two}, the attacker may try to hide only the attack information of the FDIA attacked meters (represented by the nonzero items/meters of $\boldsymbol{a}$), and the NAL's prediction results of other meters are out of his concern. If the attacker is strong enough, he will be committed to ensuring that the NAL's prediction results of all meters are normal, which is more difficult. In this case, hiding all labels can ensure that both the BDD and NAL detection modules of the power grid cannot find any abnormal phenomenon, which has great attraction and value for attackers. 

Based on the above analysis, we combine the attack objectives of the two dimensions to obtain four typical LESSON attack types as in Table \ref{type}. Obviously, LESSON-1 is the simplest type, and LESSON-4 is the most difficult type for attackers. Besides, the difficulty comparison of LESSON-2 and LESSON-3 cannot be visually determined, and will be compared and  analyzed in subsequent experiments.

\begin{table*}[htbp]
	\caption{The typical LESSON attack types based on attack objectives in two dimensions.}
	\label{type}
	\centering
	\resizebox{2.05\columnwidth}{!}{
	\begin{tabular}{|c|c|c|}
	\hline
	\backslashbox[60mm]{Objective One}{Objective Two}  & \begin{tabular}[c]{@{}c@{}}\circled{1}\\ The original FDIA attack meters are required to \\ predict ``normal" by NAL  (nonzero items/variables of $\boldsymbol{a}$)  \end{tabular} & \begin{tabular}[c]{@{}c@{}}\circled{2}\\ \begin{tabular}[c]{@{}c@{}}All the meters are predicted\\ ``normal" by NAL \end{tabular}  \end{tabular} \\ \hline
	\begin{tabular}[c]{@{}c@{}}\circled{1} The original induced estimation error (nonzero items/variables of $\boldsymbol{c}$)\\  is not required to be unchanged  \end{tabular} & LESSON-1     & LESSON-2     \\ \hline
	\begin{tabular}[c]{@{}c@{}}\circled{2} The original induced estimation error (nonzero items/variables of $\boldsymbol{c}$)\\ is required  to be unchanged \end{tabular} & LESSON-3     & LESSON-4     \\ \hline
	\end{tabular}
	}
\end{table*}

Note that for the \textbf{targeted} LESSON attack, we need to constrain and limit the optimized variables. Specifically, let $\mathbf{I}$ denote a column vector of the same size as $\boldsymbol{c}$, and the elements of $\boldsymbol{c}$ are defined by
\begin{equation}
\mathbf{I}_i=\begin{cases}
0,&		if\ \boldsymbol{c}_i\ne0\\
1,&		if\ \boldsymbol{c}_i= 0\\
\end{cases}
\end{equation}
Then the attacker can explore the corresponding \textbf{targeted} LESSON attack by solving 
\begin{subequations}
	\label{uuiooa9}
	\begin{align} 
	\label{bbbboa9}
	\min_{\boldsymbol{\varpi}} \quad  \sum_{i=1}^A{\max \left( 0, -\varUpsilon(\boldsymbol{z}_f)_j \right)}&+\sum_{j=1}^B{\max \left( 0, \varUpsilon(\boldsymbol{z}_f)_j \right)}  \\
	s.t.\ \ \ \boldsymbol{z}_f= \boldsymbol{z}_{\boldsymbol{a}}+\boldsymbol{H} & \left(\mathbf{I} \odot (\mu \cdot \tanh(\boldsymbol{\varpi}))\right)
	\end{align}	
\end{subequations}
where $\odot$ denotes the Hadamard product. In this way, we can optimize over zero items/variables in $\boldsymbol{c}$ to explore \textbf{targeted} LESSON adversarial perturbations. Of course, if we need to perturb specific state variables, we can simply assign the corresponding values (0 or 1) to the $\mathbf{I}$ vector.

\section{Experimental Analyses}
In this section, we analyze the proposed LESSON attack framework through substantial simulations with IEEE test systems, including the IEEE 14-bus, 30-bus, and 118-bus systems. Due to the page limits, the related system parameters are omitted here but can be derived from Matpower \cite{Zimmerman}.

\subsection{Experimental Setup}
\textbf{Dataset Generation:} 
Matpower \cite{Zimmerman} contains information about test systems, including their topology, bus data, and branch data, which can be used to create measurement samples. As in \cite{8660426} and \cite{esmalifalak2014detecting}, loads on each bus follow the Uniform distribution $U(80\% * base  load, 120\% * base load)$. Then, we generated 30,000 normal measurement samples under the assumption that the test systems are thoroughly measured. 15,000 samples are utilized as attacked/false training instances when well-designed FDIA vectors are added, and 15,000 samples are utilized as legitimate/normal training data. The noise added has a zero-mean Gaussian distribution, and the standard deviation is set at 2$\%$ of the mean measurements of the relevant meter. The FDIA vectors ($\boldsymbol{a}=\boldsymbol{Hc}$) were created based on randomly generated $\boldsymbol{c}$ \cite{9622117}: 

(i) The number of targeted state variables follows the Uniform distribution $U(1,n/2$); 

(ii) The targeted state variables follow the Gaussian distribution $N\sim \left( 0,\nu ^2 \right)$, where $\nu ^2$ represents the variance. 

Then, according to variance differences, we created 5000 FDIA vectors at each of the following three scales:

(i) \textbf{small scale}, $\nu ^2=0.02$;\

(ii) \textbf{medium scale}, $\nu ^2=0.1$;

(iii) \textbf{large scale}, $\nu ^2=0.5$.

After the above process, we eventually obtained 15,000 normal and 15,000 attacked data by adding these 15,000 well-designed FDIA vectors at random to samples of 15,000 of 30,000 normal instances. The entire dataset was then randomly divided at a 2:1 ratio, with 20,000 samples for training and 10,000 samples for testing.

\textbf{Target CNNs for NAL:} Since CNN (Convolutional Neural Network) can capture the inconsistency and cooccurrence dependency introduced by FDIA and its excellent locational detection performance \cite{9049087,tiis:24681, Mukherjee2022,MUKHERJEE2022100702,en15145312,Qin2022}, the NAL models for FDIA locational detection were created using CNN. The network architectures of CNN models for NAL are given in Table \ref{CNN model}, and Pytorch is used to train the corresponding NAL models. The representative well-trained NAL models' test meter accuracies (row accuracies \cite{9049087}) for the 14-bus, 30-bus, and 118-bus systems, are 99.6$\%$ (97.6$\%$), 99.4$\%$ (96.2$\%$), and 99.1$\%$ (93.1$\%$), respectively.

\begin{table*}[htbp]
	\caption{Network architectures of CNN models used to implement NAL methods in experimental analyses for IEEE 14-bus, 30-bus and 118-bus systems. Conv:Convolution, BN:BatchNorm, LR:LeakyReLU, FL:Flatten, FC:FullyConnect, SM:Sigmoid.}
	\label{CNN model}
	\centering
	\resizebox{2.05\columnwidth}{!}{
	\begin{tabular}{|c|c|c|c|c|c|c|c|c|c|c|c|c|c|c|c|c|c|c|c|c|c|}
	\hline
	CNN models & 1                                                     & 2  & 3  & 4                                                    & 5  & 6  & 7                                                    & 8  & 9  & 10                                                   & 11 & 12 & 13                                                   & 14 & 15 & 16                                                   & 17 & 18 & 19 & 20 & 21 \\ \hline
	14-bus     & \begin{tabular}[c]{@{}c@{}}Conv\\ (10*1)\end{tabular} & BN & LR & \begin{tabular}[c]{@{}c@{}}Conv\\ (5*1)\end{tabular} & BN & LR & \begin{tabular}[c]{@{}c@{}}Conv\\ (3*1)\end{tabular} & BN & LR & \begin{tabular}[c]{@{}c@{}}Conv\\ (3*1)\end{tabular} & BN & LR & FL                                                   & FC & SM &                                                      &    &    &    &    &    \\ \hline
	30-bus     & \begin{tabular}[c]{@{}c@{}}Conv\\ (10*1)\end{tabular} & BN & LR & \begin{tabular}[c]{@{}c@{}}Conv\\ (5*1)\end{tabular} & BN & LR & \begin{tabular}[c]{@{}c@{}}Conv\\ (3*1)\end{tabular} & BN & LR & \begin{tabular}[c]{@{}c@{}}Conv\\ (3*1)\end{tabular} & BN & LR & \begin{tabular}[c]{@{}c@{}}Conv\\ (3*1)\end{tabular} & BN & LR & FL                                                   & FC & SM &    &    &    \\ \hline
	118-bus    & \begin{tabular}[c]{@{}c@{}}Conv\\ (5*1)\end{tabular}  & BN & LR & \begin{tabular}[c]{@{}c@{}}Conv\\ (5*1)\end{tabular} & BN & LR & \begin{tabular}[c]{@{}c@{}}Conv\\ (5*1)\end{tabular} & BN & LR & \begin{tabular}[c]{@{}c@{}}Conv\\ (3*1)\end{tabular} & BN & LR & \begin{tabular}[c]{@{}c@{}}Conv\\ (3*1)\end{tabular} & BN & LR & \begin{tabular}[c]{@{}c@{}}Conv\\ (3*1)\end{tabular} & BN & LR & FL & FC & SM \\ \hline
	\end{tabular}
	}
\end{table*}

\textbf{ Evaluation of Attacks:} Since the NAL models' test accuracies are less than 100$\%$, we only select samples that are completely correctly predicted\footnote{The completely correctly predicted sample means that all the uncompromised meters are labeled as normal and all the compromised meters are labeled as attacked.} to carry out the proposed LESSON attacks. Specifically, for each power system and each FDIA attack scale, 500 completely correctly predicted samples are randomly selected, and the proposed four LESSON attacks are employed to generate multi-label adversarial perturbations, where $\mu$ is set as $1 \ rad = (180/\pi) ^o$ as in \cite{9695995}, and the maximum iteration number of Adam is set as 500 (the algorithm will automatically terminate when a suitable adversarial perturbation is found). Since the success rates ($P_{suc}$\footnote{For a successful case, the suitable adversarial perturbation satisfying the attack objectives in two dimensions is found.}) can vary depending on the circumstances, we examined the attack success rate $P_{suc}$ and impacting factors in various scenarios. Besides, we also evaluate the magnitude of adversarial perturbations in successful attack samples using the following two metrics:

\begin{equation}
	\label{pert_c}
	\rho_{\boldsymbol{c}}=\frac1{|D|}\sum_{\boldsymbol{x}\in D}{\|\boldsymbol{\zeta}\|_2},
\end{equation}
\begin{equation}
	\label{pert_a}
	\rho_{\boldsymbol{a}}=\frac1{|D|}\sum_{\boldsymbol{x}\in D}{\|\boldsymbol{H} \boldsymbol{\zeta}\|_2},
\end{equation}
where $\boldsymbol{\zeta}$ represents the perturbation added to state variables, $\boldsymbol{H} \boldsymbol{\zeta}$ is the final adversarial perturbation for the FDIA measurement vector $\boldsymbol{z}_{\boldsymbol{a}}$, and $D$ represents the dataset composed of successful attack samples.

\subsection{Analysis of Attack Objectives in Two Dimensions}
~\

\begin{figure*}[htbp]
	\centerline{\includegraphics[scale=0.36]{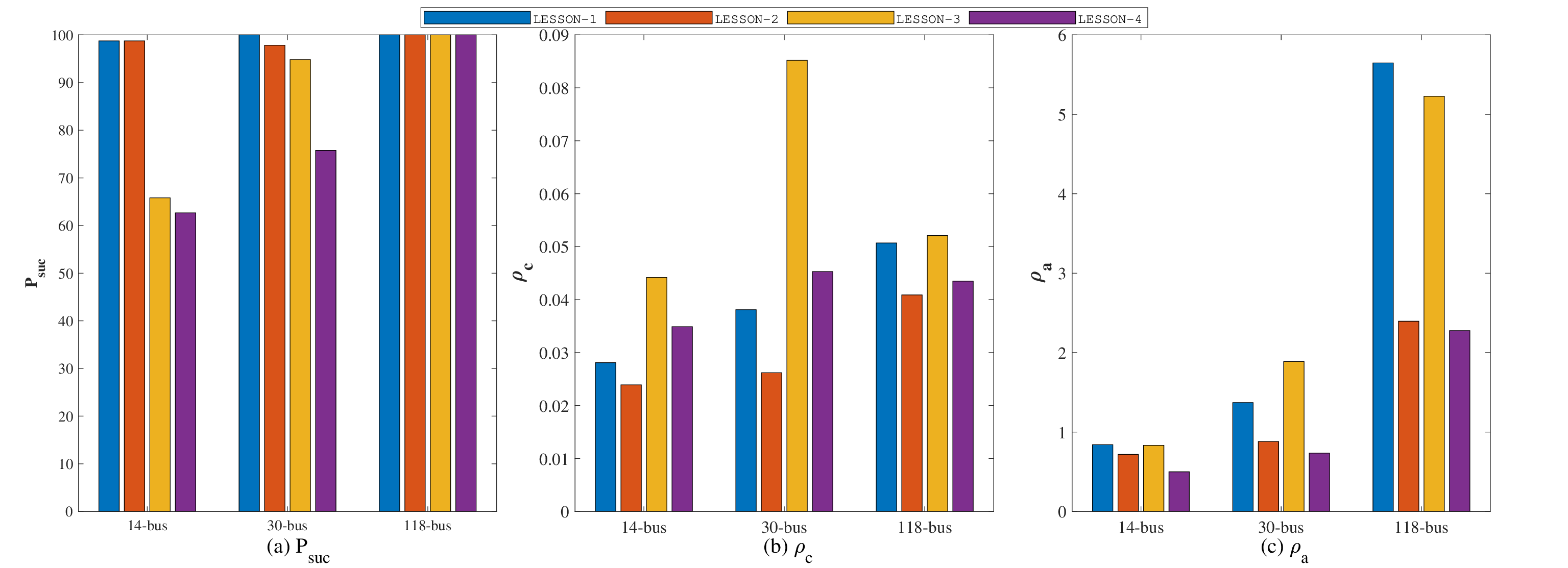}}
	\centering
	\caption{$P_{suc}$, $\rho_{\boldsymbol{c}}$, and $\rho_{\boldsymbol{a}}$ with four different LESSON attacks: the original learning rate $\alpha$ is 0.001, and the original FDIA attacks are at the small scale.}
	\label{attack_objectives}
\end{figure*}

We first explored the proposed four typical LESSON attacks against small-scale FDIA cases, where the initial learning rate of Adam is set as 0.001. The related results presented in Fig.~\ref{attack_objectives} (a) show the excellent attack success rate of the proposed LESSON framework. The most difficult LESSON-4 case of the 14-bus and 30-bus systems still has a success rate of more than 60$\%$, and the success rate of most other LESSON cases is close to 100$\%$, which means that most LESSON attacks can successfully achieve the intended attack objectives. Besides, the success rate of all four LESSON attacks of the 118-bus system is 100$\%$, which is much higher than that of the 14-bus and 30-bus systems, especially in the more difficult cases LESSON-3 and LESSON-4. Through the comparative analysis of these four LESSON attacks, we find that \textbf{Objective One} (if the original induced estimation error need to remain unchanged or not) has a greater impact on the attack success rate. Specifically, keeping the original induced estimation error unchanged (\textbf{targeted} LESSON) obviously increases the difficulty of attack. On the contrary, the impact of \textbf{Objective Two} is relatively small, although it also affects the attack success rate. Note that, LESSON-2 and LESSON-4 mean hiding all the labels of meters, while LESSON-4 is also committed to achieving predetermined estimation errors. Besides, LESSON-3 is to ensure that the FDIA attacked meters are not detected to achieve the predetermined estimation error, which may cause NAL to identify normal meters as attacked. The above situations and their high attack success rates show that the proposed LESSON attack poses a huge threat to large-scale power systems. In addition, $\rho_{\boldsymbol{c}}$ and $\rho_{\boldsymbol{a}}$ are presented in Fig.~\ref{attack_objectives} (b) and (c), respectively. From Fig.~\ref{attack_objectives} (b), one can observe that LESSON-1 needs larger adversarial perturbations than LESSON-2, and LESSON-3 needs larger adversarial perturbations than LESSON-4. We believe that the reason behind this is that LESSON-1 and LESSON-3 are easier attacks because only the original FDIA attack meters are required to  predict normal, and the proposed attack framework and corresponding iterative algorithm are easier to find suitable adversarial perturbations without subsequent perturbation optimization, resulting in relatively large adversarial perturbations. On the contrary, LESSON-1 needs smaller adversarial perturbations than LESSON-3, and LESSON-2 needs smaller adversarial perturbations than LESSON-4, although the former's attacks are easier to achieve because the latter also needs to ensure that the original induced estimation error remains unchanged. We speculate that it is the condition of ensuring the original induced estimation error remains unchanged that leads to larger adversarial perturbations. If the original induced estimation error can be changed, reducing its magnitude and adding smaller perturbations to other state variables can also help avoid detection. However, if it is necessary to ensure that the original induced estimation error remains unchanged, then relatively larger adversarial perturbations need to be added to other state variables to avoid detection. It is worth noting that the trend and pattern of changes in Fig.~\ref{attack_objectives} (b) and (c) are not the same, that is, the larger type in Fig.~\ref{attack_objectives} (b) may not necessarily be larger in Fig.~\ref{attack_objectives} (c). The reason is that there is no strict proportional relationship between $\|\boldsymbol{\zeta}\|_2$ and $\|\boldsymbol{H} \boldsymbol{\zeta}\|_2$. However, based on the LESSON attack framework and appropriate parameter settings, the resulting adversarial perturbations including $\|\boldsymbol{\zeta}\|_2$ and $\|\boldsymbol{H} \boldsymbol{\zeta}\|_2$ are both small, which means the LESSON attack framework is effective.

\subsection{Analysis of Effect of FDIA Scales}
\begin{figure*}[htbp]
	\centerline{\includegraphics[scale=0.36]{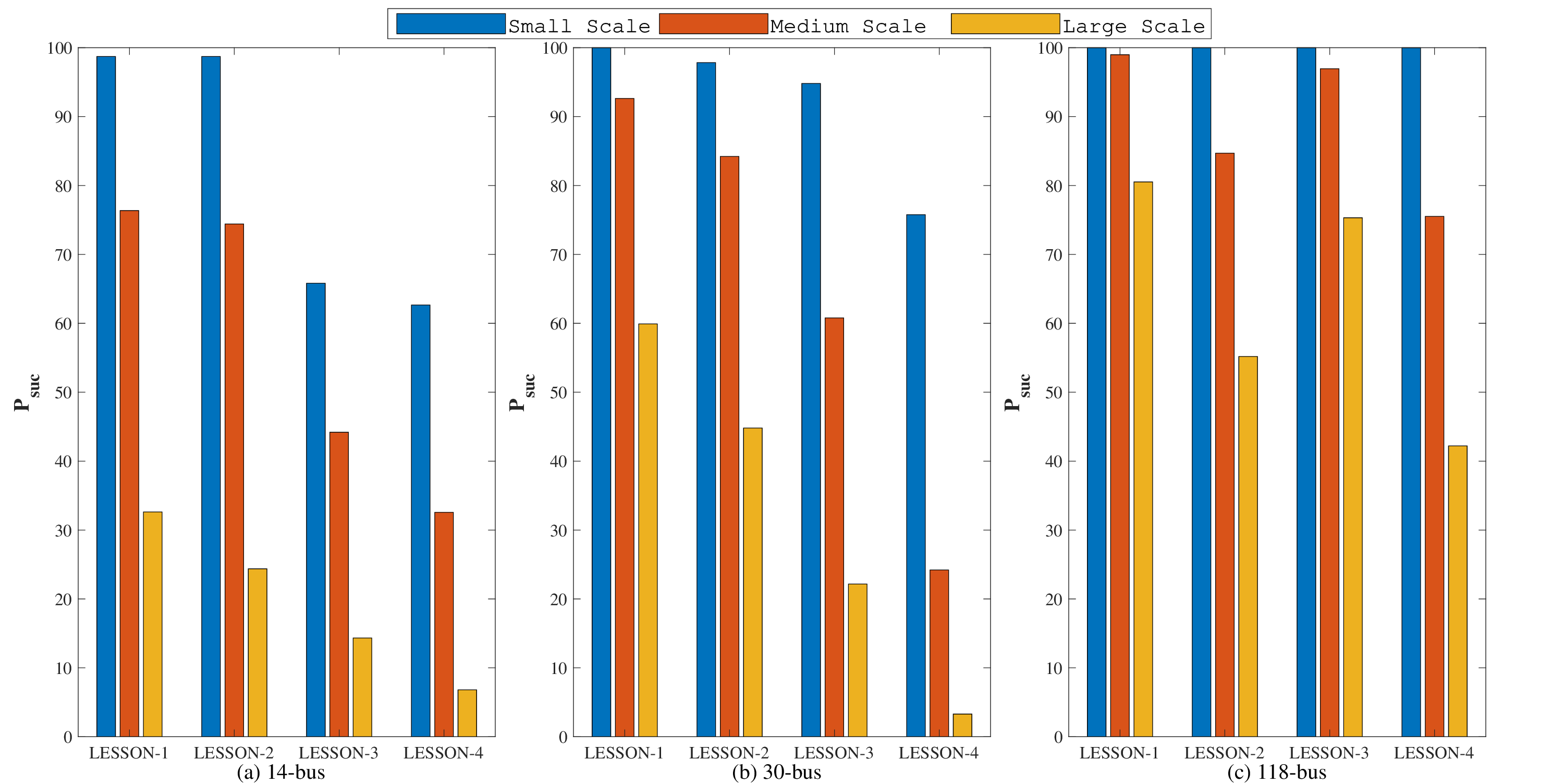}}
	\centering
	\caption{$P_{suc}$ with three different FDIA attack scales: the original learning rate is 0.001.}
	\label{attack_scale}
\end{figure*}

\begin{figure*}[htbp]
	\centerline{\includegraphics[scale=0.36]{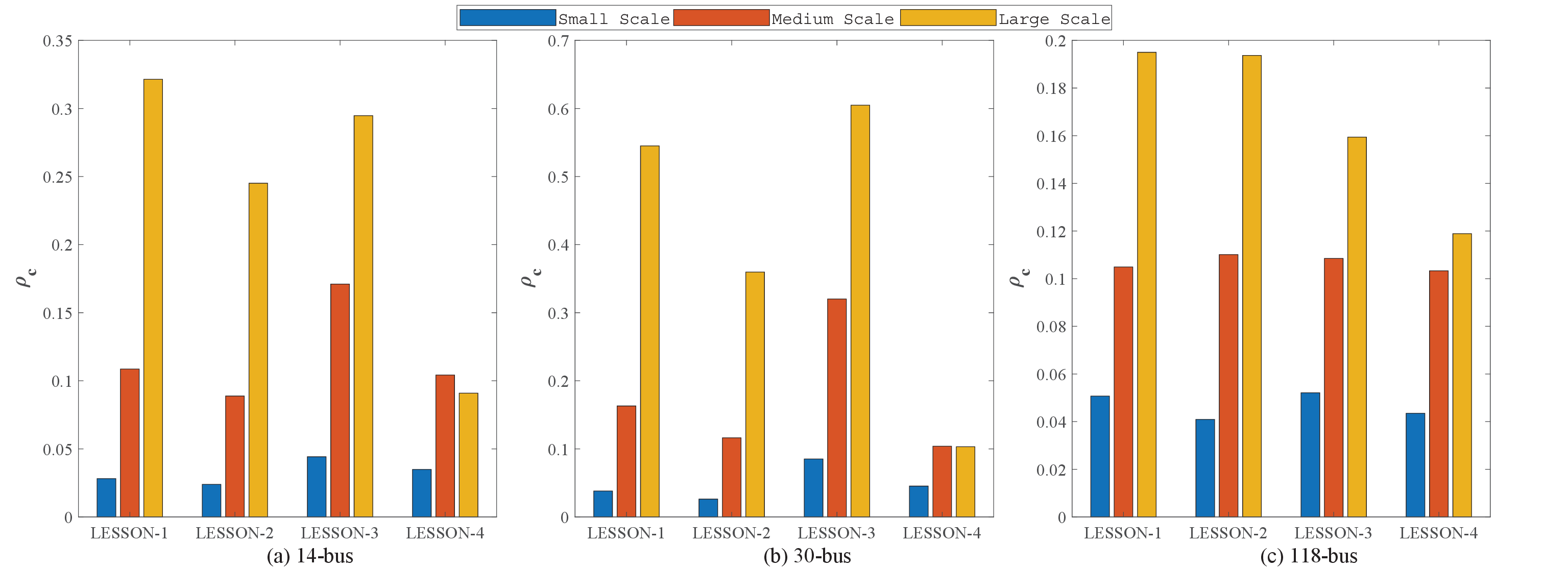}}
	\centering
	\caption{$\rho_{\boldsymbol{c}}$ with three different FDIA attack scales: the original learning rate is 0.001.}
	\label{attack_scale_2}
\end{figure*}

We then investigated the proposed four typical LESSON attacks against different FDIA scales, and the initial learning rate of Adam is also set as 0.001. The related results are shown in Fig.~\ref{attack_scale}, which indicate that the FDIA attack scale does have a significant influence on the attack success rate. Take the LESSON-2 of 14-bus system for example (Fig.~\ref{attack_scale} (a)), as the attack scale increases, the attack success rates are 98.73$\%$, 74.42$\%$, and 24.37$\%$, respectively, showing a significant decline. We speculate that if the FDIA attack scale is relatively large, the adversarial perturbation required to bypass the NAL's detection may also be relatively large. However, the suitable adversarial perturbation is limited by the physical constraints, which leads to a decline in the attack success rate. By comparing the three power systems, we find that with the increase of power grid scale, the attack success rate is also increasing, especially for scenarios with large FDIA attack scale. Specifically, for the 118-bus system, the success rates for LESSON-1 to LESSON-4 with large FDIA attack scale are 80.52$\%$, 55.19$\%$,75.32$\%$, and 42.21$\%$, respectively. This phenomenon once again presents the vulnerability of large-scale power systems to the proposed LESSON attacks. In addition, the results presented in Fig.~\ref{attack_scale_2} indicate that as the FDIA attack scale increases, the added adversarial perturbation also needs to be larger to bypass detection, which is consistent with our intuitive idea. Although the large scale scenarios of LESSON-4 in 14-bus and 30-bus do not fully comply with the above rules, this is due to the extremely low attack success rate and limited successful attack samples at this time (see Fig.~\ref{attack_scale}), while other scenarios are consistent with the above patterns and characteristics.

\subsection{Analysis of the Learning Rate of Adam}
\begin{figure}[htbp]
	\centerline{\includegraphics[scale=0.57]{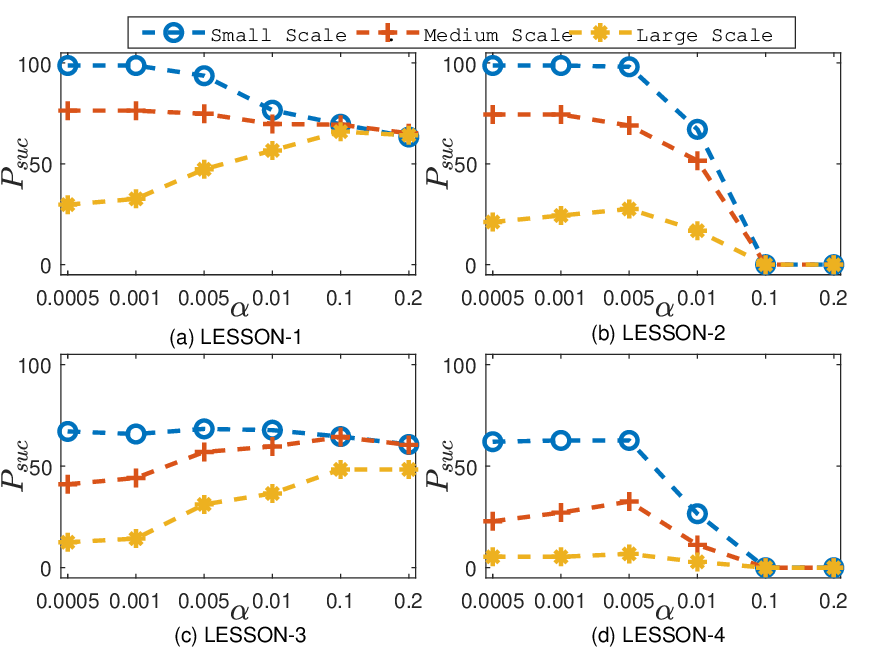}}
	\centering
	\caption{The success rates $P_{suc}$ for IEEE 14-bus system with different Adam's original learning rates.}
	\label{14bus_adam}
\end{figure}

\begin{figure}[htbp]
	\centerline{\includegraphics[scale=0.57]{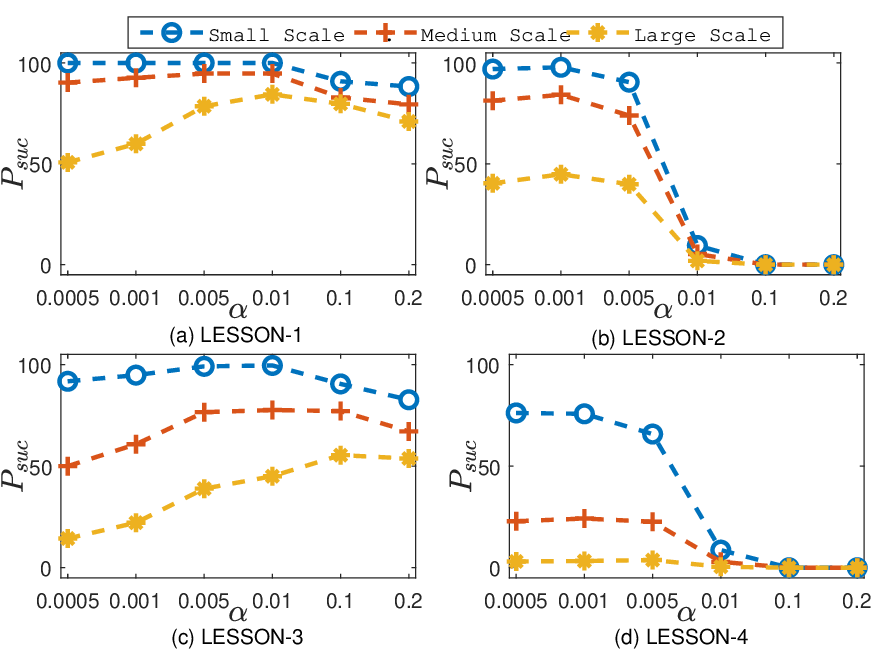}}
	\centering
	\caption{The success rates $P_{suc}$ for IEEE 30-bus system with different Adam's original learning rates.}
	\label{30bus_adam}
\end{figure}

\begin{figure}[htbp]
	\centerline{\includegraphics[scale=0.57]{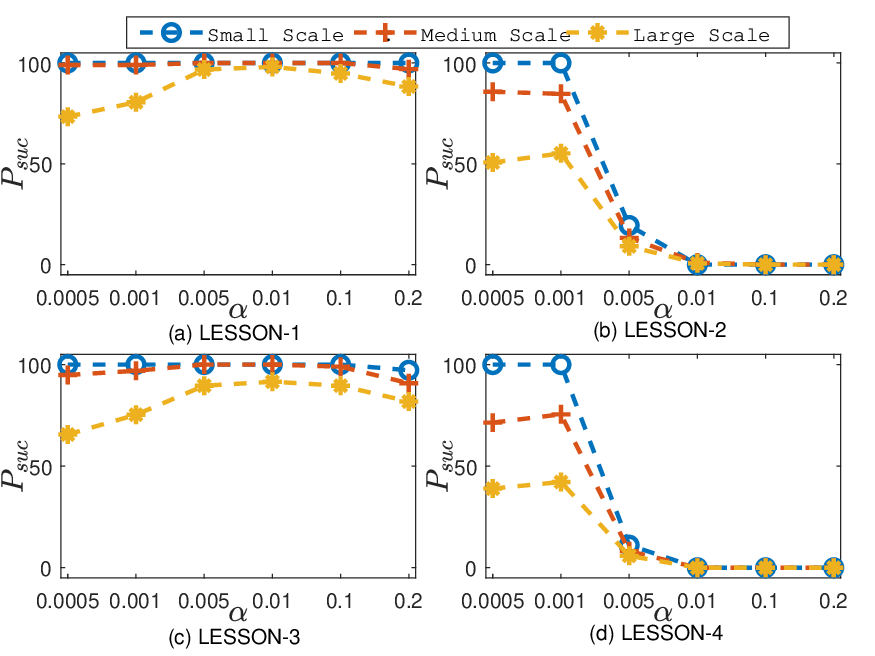}}
	\centering
	\caption{The success rates $P_{suc}$ for IEEE 118-bus system with different Adam's original learning rates.}
	\label{118bus_adam}
\end{figure}

Since the Adam algorithm is employed to generate multi-label adversarial perturbations, we then explored the effect of the initial learning rate of Adam on the attack success rate. Multiple learning rates (0.0005, 0.001, 0.005, 0.01, 0.1, 0.2) are examined (the other parameters of Adam are set to default values because their impact is relatively small), and the corresponding results are in Fig.~\ref{14bus_adam}, \ref{30bus_adam}, and \ref{118bus_adam}, respectively. Since the influence of the learning rate is related to the FDIA attack scale, we analyze the situation of different FDIA attack scales separately. 
\begin{enumerate}
	 \item \textbf{Small-Scale}: the success rate of almost all power systems decreases with the increase of the original learning rate. The reason may be that for small-scale FDIA attacks, a smaller learning rate can produce more subtle perturbations, which will make it easier to find a suitable adversarial perturbation. For LESSON-1 and LESSSON-3, the decrease of success rate caused by the increase of the original learning rate is not obvious, and it is relatively obvious only after the learning rate reaches 0.1 or about 0.01. In contrast, the success rate of both LESSON-2 and LESSSON-4 declined significantly and began to decline significantly when the learning rate just reached 0.001. 
	 \item \textbf{Medium-Scale}: with the increase of the original learning rate, the success rate for LESSON-3 and LESSON-4 increases first and then decreases. The point where the success rate changes from increase to decrease is about 0.001. For LESSON-1, the decrease of success rate caused by the increase of the original learning rate is not obvious, and it is relatively obvious only after the original learning rate reaches 0.1. On the contrary, the success rate of LESSON-2 began to decline significantly when the original learning rate just reached 0.001. 
	 \item \textbf{Large-Scale}: for each system and each LESSON attack, with the increase of the original learning rate, the success rate first increases and then decreases. We speculate that for large-scale FDIA vectors, under the limited number of iterations, an appropriately large initial learning rate is helpful to improve the efficiency of finding the adversarial perturbation, thus increasing the attack success rate. For LESSON-1 and LESSON-3, the success rates of all power systems are almost always rising, and most of them begin to decline only after the original learning rate reaches 0.1 or 0.01. For LESSON-2 and LESSON-4, the success rates of all power systems rise slightly and then drop rapidly to 0.
\end{enumerate} 

Through the above analysis, we can conclude the following characteristics: with the increase of the FDIA attack scale, the initial learning rate of Adam should also be appropriately increased to seek appropriate adversarial perturbations more efficiently, thus improving the attack success rate. According to the above experimental results, the attack success rate is related to the attack scale and attack target: for the case \textbf{Objective Two} \circled{2} (LESSON-2 and LESSSON-4), the original learning rate should be somewhat smaller, otherwise the success rate will quickly decrease to 0; for other cases, the chosen original learning rate cannot be too large or too small, which can be roughly in the range of [0.001,0.1] and adjusted according to the FDIA attack scale. 

\subsection{Analysis of Effect of State Perturbation Range}
\begin{figure}[htbp]
	\centerline{\includegraphics[scale=0.57]{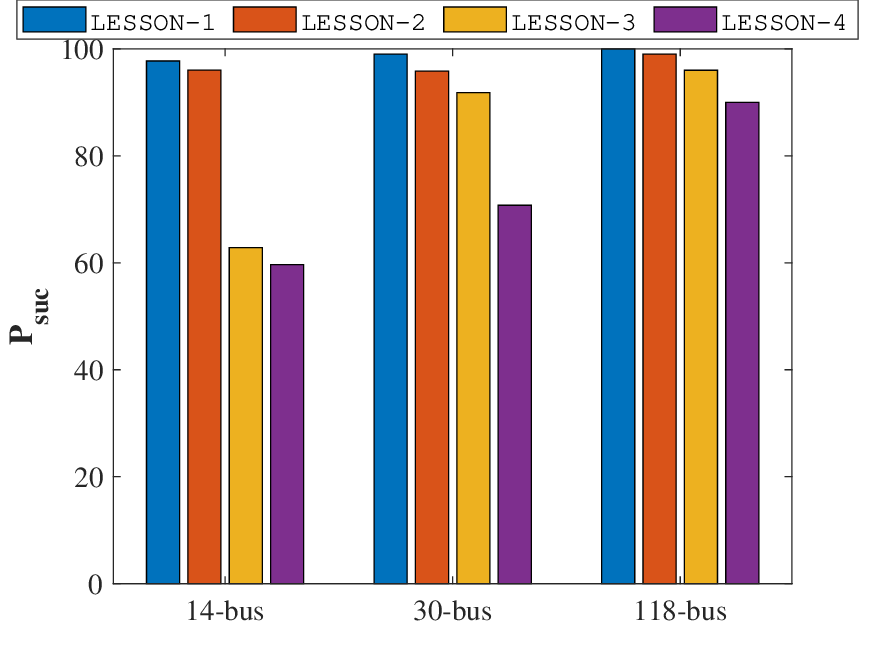}}
	\centering
	\caption{The success rates $P_{suc}$ with four different LESSON attacks: $\mu$ is set as $0.5 \ rad$, the original learning rate $\alpha$ is 0.001, and the original FDIA attacks are at the small scale.}
	\label{attack_objectives_mu0.5}
\end{figure}

We then investigated the impact of state perturbation range for the proposed LESSON attack framework. $\mu$ is set as $0.5 \ rad = (90/\pi) ^o$, and the related results are shown in Fig.~\ref{attack_objectives_mu0.5}. Comparing Fig.~\ref{attack_objectives} and Fig.~\ref{attack_objectives_mu0.5}, we can see that reducing the state perturbation rage from $1 \ rad$ to $0.5 \ rad$ has some minor impact on the attack success rate. Due to the state perturbation range affecting the scope and space of adversarial perturbations that the LESSON attack framework seeks, we can intuitively speculate that if we continue to lower the state perturbation range, the attack success rate will continue to decrease. For example, if $\mu$ is set as a very small value such as  $0.01 \ rad$, then the attack success rate in many cases will significantly decrease to approximately 0. However, we believe that such a setting ($1 \ rad$ or $0.5 \ rad$) satisfies the constraints and normal laws of power systems, and is also sufficient to find suitable adversarial perturbations.

\subsection{Extending to other LESSON attack types}
Although we only evaluated the four typical LESSON attack types in the experiment, the proposed LESSON framework can be extended to other attack types. For example, in the process of generating adversarial perturbations, attack cost limitations should be considered. Due to the fact that the final LESSON attack not only bypasses BDD detection, but also NAL's detection, additional attack costs (controlling additional meters to add adversarial perturbations) are necessary unless the attacker considers changing the original attack objective, such as reducing the magnitude of the initial attack objective. However, in practical situations, the cost of attack is always limited. Assuming that there are some meters that the attacker cannot control even after increasing a certain attack cost, then the attacker cannot tamper with the data of these meters, which can be added to the LESSON framework and corresponding models. Taking the mathematical model (\ref{uuiooa7}) for example, we can make corresponding modifications:
\begin{subequations}
	\label{uuiooa72}
	\begin{align} 
	\label{bbbboa72}
	\min_{\boldsymbol{\varpi}} \quad  \sum_{i=1}^A{\max \left( 0, -\varUpsilon(\boldsymbol{z}_f)_j \right)}&+\sum_{j=1}^B{\max \left( 0, \varUpsilon(\boldsymbol{z}_f)_j \right)}  \\
	s.t.\ \ \ \boldsymbol{z}_f= \boldsymbol{z}_{\boldsymbol{a}}+&\boldsymbol{H} (\mu \cdot \tanh(\boldsymbol{\varpi}))\\
	\label{cccc2}
	(\boldsymbol{z}_f)&_i=0, i\in N,
	\end{align}	
\end{subequations}
where $N$ represents the meter set without control, and $\mu$ represents the predetermined state perturbation range. The added constraint condition (\ref{cccc2}) can be processed during the iteration process: in each iteration optimization of adversarial perturbation, the perturbation values of meters in $N$ are set to 0, and subsequent iterative optimization is carried out. Similar methods and techniques have been widely used in projected gradient descent (PGD) related adversarial attacks in the image field \cite{DBLP}, which are very mature.

\begin{figure}[htbp]
	\centerline{\includegraphics[scale=0.57]{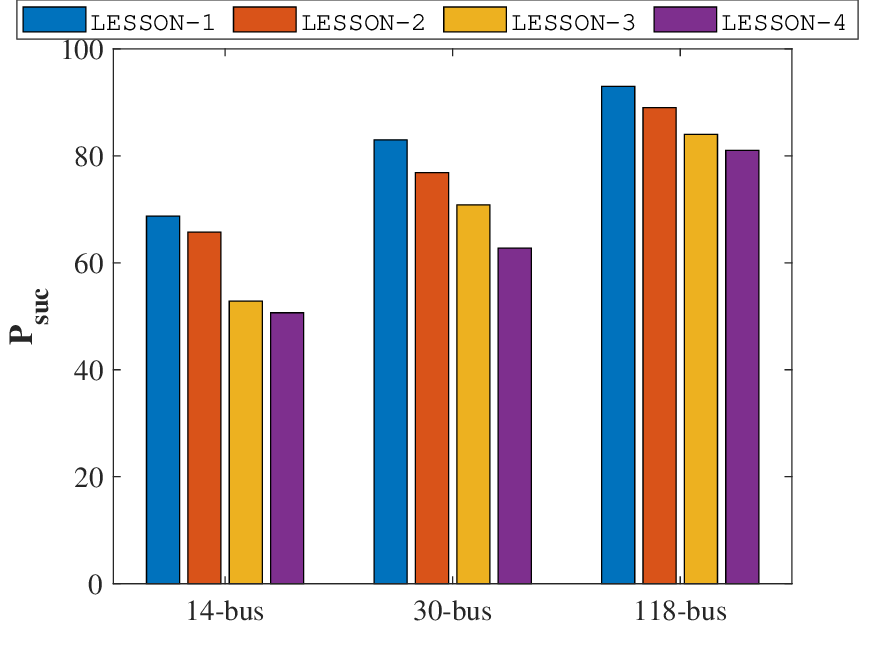}}
	\centering
	\caption{The average success rates $P_{suc}$ for the mathematical model (\ref{uuiooa7}): $\mu$ is set as $1 \ rad$, the original learning rate $\alpha$ is 0.001, and the original FDIA attacks are at the small scale.}
	\label{attack_cost_half}
\end{figure}

In order to test the effectiveness of the above attack type, we randomly set half of the meters, except for those that must be controlled (corresponding to non-zero items/meters of $\boldsymbol{a}$ by $\boldsymbol{a}=\boldsymbol{Hc}$), as the meter set without control. Then, we use the above model and corresponding iterative algorithm to solve for adversarial perturbations and record the average success rate $P_{suc}$, as shown in Fig.~\ref{attack_cost_half}. Comparing Fig.~\ref{attack_cost_half} and Fig.~\ref{attack_objectives} (a), we can find that the reduction of attack costs will reduce the success rate of attacks, which is consistent with our intuitive understanding. Even so, the proposed LESSON framework and iterative algorithm can still find suitable adversarial perturbations under limited conditions and maintain a certain attack success rate. In addition, the proposed LESSON framework can also be extended according to other actual situations. Of course, as the restrictions on attack conditions increase, the attack success rate will decrease, which is also very normal and intuitive.

\section{Conclusion}
In this work, we designed the LESSON framework to explore multi-label adversarial false data injection attacks. The proposed LESSON attack framework includes three key designs, namely Perturbing State Variables,  Tailored Loss Function Design, and Change of Variables, which can help find suitable multi-label adversarial perturbations within the physical constraints to avoid both BDD and NAL's detection. We explored four typical LESSON attacks (LESSON-1 to LESSON-4) based on the proposed attack framework and two dimensions of attack objectives and conducted extensive experimental analyses to investigate the related four LESSSON attacks. The experimental results demonstrate that the proposed LEESON attack framework has a very good attack success rate, which poses a serious and imperative security breach and risk for practical large-scale power systems.

However, this study may have some potential shortcomings. First, the premise of the white-box LESSSON attack may not always be accurate. However, due to APT's (Advanced Persistent Threat) well-funded, experienced teams of adversaries, white-box attacks against safety-critical power systems cannot be ignored. In this case, the white-box attack is employed to investigate and evaluate the reliability and weaknesses of multi-label FDIA locational detectors. Many cyber security  studies also begin with potential white-box attacks \cite{goodfellow2014explaining} and then turn to more practical black-box attacks \cite{papernot2017practical}, like the adversarial examples in the images domain. Therefore, in our future work, we will investigate black-box multi-label adversarial false data injection attacks. Besides, the assumption of attack cost in the attack model is ideal, which means that the attacker has rich attack resources. In the case of very limited attack resources, whether the attacker can implement such LESSON attacks or whether multi-label FDIA locational detectors are safe enough is a subject worthy of further comprehensive research. Although we conducted preliminary analysis in the experimental analysis section, we need to explore and analyze this problem in the follow-up work.

Second, although the success rate of attacks increases as the scale of the power grid increases in some cases, we do not have conclusive experimental evidence to prove the relationship between attack success rate and the size of the power grid. It is worth noting that the scalability of attacks in large-scale power systems is complex but also meaningful. Therefore, the scalability of attacks in large-scale power systems deserves the attention of relevant researchers and further research in the future.

Third, the effective defense methods against multi-label adversarial examples have not been studied yet. The common defense measures for single-label adversarial examples, such as adversarial training \cite{kurakin2016adversarial}\cite{tramer2017ensemble}, defensive distillation \cite{papernot2016distillation}, and adversarial detection, may not be fully suitable for multi-label application scenarios \cite{JinyuanJia}. The related characteristics of multi-label learning \cite{9661418} and domain knowledge of power systems may provide provide strong support for the design of effective defense measures, which will be studied in the future.


%




\ifCLASSOPTIONcaptionsoff
  \newpage
\fi



%
\bibliographystyle{IEEEtran}
\bibliography{IEEEabrv,mybibfile}

\end{document}